LETTER TO THE EDITOR

# AC conductivity of polymer composites: an efficient confirmatory tool for qualifying crude multi-walled carbon nanotube-samples

**Soumen Karmakar[1], Harshada Nagar[1], J P Jog[2], S V Bhoraskar[1,4] and A K Das[3]**

[1]Department of Physics, University of Pune, Pune 411007, India
[2]National Chemical Laboratory, Pashan Road, Pune 411008, India
[3]Laser and Plasma Technology Division, Bhabha Atomic Research Centre, Mumbai 400085, India
[4]Author to whom any correspondence should be addressed
E-mail: svb@physics.unipune.ernet.in

**Abstract**
The present communication highlights that ac conductivity measurement of the multi-walled carbon nanotubes (MWNTs)-polymer composites is a very promising characterizing tool for qualifying any crude MWNT-sample synthesized by electric arc. It distinguishes graphite structures from that of the MWNTs and reflects the relative amount of nearly one-dimensional structures and disorders present within the samples. This new protocol would help in evolving more direct and quantitative criteria for qualification of MWNTs synthesized through diverse techniques and will definitely add up to the conceptual understanding in visualizing the exact roles of different controlling parameters affecting the growth of MWNTs.

Since the advent of multi-walled carbon nanotubes (MWNTs) [1], this novel exotic material has been one of the most important researched themes all across the world [2]. MWNTs related studies encompass not only understanding of their formation mechanisms [3] and strange properties, but have also been directed towards exploring various synthesis routes as well as applications [4, 5].

Arc plasma method being operated above the graphite sublimation point for the synthesis of MWNTs provides very high temperature and enthalpy and is known to yield the best quality MWNTs belonging to the fullerene family [6]. Catalytic chemical vapour deposition techniques and pyrolytic processes, though being promoted most widely for the bulk production of MWNTs now-a-days, have a probability of co-producing both MWNTs and carbon nano fibers (CNFs) [7], thereby hindering the actual MWNT-properties to be examined in bulk and employing them further in various application areas. On the contrary, arc plasma processes have never been known to generate CNFs, which to our understanding is very difficult to differentiate with the MWNTs envisaged through any known bulk characterizing tool.

However, in spite of being recognized as the source of pristine and well-graphitized MWNTs, arc plasma technique has rather been ignored in the recent years. This is mostly due to a huge amount of loss of the consumed material in the forms of amorphous carbon, fullerenes and carbon nanocrystalline particles [2]. Efforts directed towards the improvement of the arc-synthesized MWNTs in terms of purity and yield, are therefore, of real challenge.

Since the discovery of MWNTs, till date there are a number of publications [8-16], which claim that by varying suitable operating parameters and on account of diverse methodologies both the purity and yield of arc-derived MWNTs can be controlled. Very recently the authors have also reported about a very promising arc-technique which has been shown to generate MWNTs with orders of magnitude higher yield and enhanced purity [17]. These studies [8-17] are of extreme strategic and conceptual importance, because; the exact understanding of the entire process phenomena involved therein is definitely the key to provide the required level of understanding to introduce determinism currently lacking in this field of research. We believe that it is time to introduce deterministic approaches in order to facilitate the mass production of MWNTs in the real sense using arc, rather than waiting for a serendipitous discovery to happen.

However, for determinism to be introduced in the field of arc synthesis of MWNTs, one has to have a thorough understanding of the controlling parameters affecting the growth of MWNTs. To our belief, in order to understand the exact roles of the controlling parameters involved in a number of studies [8-16] and for relying completely on the conclusions obtained thereof, they should pass through another quality assuring filter, which is reliable and has the potential to differentiate between any two MWNT samples with increased sensitivity. This is because; the conclusions obtained using available MWNT characterizing tools may often be misleading due to the following facts recognized by the authors.

The common qualifying techniques used for estimating the quality of a MWNT-sample are transmission electron microscopy (TEM), scanning electron microscopy (SEM), scanning tunneling microscopy (STM), atomic force microscopy (AFM), Raman spectroscopy (RS), thermogravimetry (TG), x-ray diffraction (XRD) analysis, x-ray photoelectron spectroscopy, infrared spectroscopy (IRS), electron paramagnetic resonance (EPR) [18-21] etc. However, with all these characterizing tools, can one unambiguously certify the relative percentages and purities (defect content) of different batches of arc-produced crude MWNTs? The present communication has looked at finding an answer to this problem and alternatively has suggested a protocol based on a simple, reliable, easy-to-undertake and cost-effective technique, which has so far been ignored completely for characterizing different batches of arc-produced crude MWNTs.

MWNTs are recognized as equi-spaced, wrapped up graphene sheets with the nearly one dimensional structure [1]. Due to the presence of multiple graphene sheets, MWNTs very closely resemble the graphite structure [2]. Interestingly, both graphite and MWNTs give similar signatures when examined through IRS, XRD and EPR [18]. Though the signatures of MWNTs slightly differ [18] from those of the graphite, accurate identification of the two carbon structures may always be a tricky affair and therefore, may often be misleading.

However, RS and TG are realized as much superior tools which are efficient in making qualitative comparisons among different batches of MWNTs. RS may prove itself to be very authentic when the RS spectra are accompanied with the radial breathing modes (RBMs) of MWNTs. It is well known that only those MWNTs, which are devoid of any structural defect and have inner diameters less than 1nm, can exhibit RBMs [22]. From the positions of the RBM lines, predictions can be made to infer the inner-wall thickness and the number of walls of the MWNTs [23]. However, till date, RBMs of MWNTs could be observed only in very rare occasions including our previous report [13, 22-25]. Most of the MWNT-samples reported in the literature have shown only D and G bands [19] in the first order RS. However, high purity MWNTs with high degree of graphitization only give rise to G band thereby giving the RS spectra very close resemblance with that of the highly oriented pyrolytic graphite [21]. Moreover, it is really very difficult to differentiate the RS spectra originating from the crude MWNT samples, CNFs and the roughly oriented graphite, as all of them give rise to similar D and G bands [19, 26].

TG is well suited for making a comparative study of different arc-generated MWNT-samples. It is known that arc-generated MWNTs due to their outstanding graphitized nature have a much higher most probable oxidation temperature ($T_P$) than the graphite and carbon nanocrystalline particles (CNPs). For example, proper judgment can be made if one compares $T_P$ of different arc-generated samples (cathode deposits), which are known to contain only MWNTs and CNPs [17]. However, this method has also its own limitations. The fundamentals of this technique prohibit it to exclusively recognize MWNTs from other carbon allotropes and the conclusions sought out solely from the TG data may often be misleading because of overlapping signatures of CNPs and structural defects present within an arc-generated MWNT-sample.

On the other hand, sophisticated tools like TEM, SEM, STM, AFM and SPM provide two-dimensional features of the three-dimensional species with varying sizes, shapes and mass density. The results depend significantly on the diversity in the preparation recipe of the samples [27]. Moreover, these microscopic data and the features observed thereof are not exactly reproducible even over repeated scans of the same sample and thus, the doubt whether these features, observed over the scanned area, do really have an one to one correspondence with the overall quality of the MWNT-samples or not, cannot be hooted out from a realistic and practical point of view. The most commonly used TEM analysis may also provide wrong information about the MWNT-samples because of the use of support films, which is reported to contain a variety of carbon structures including MWNTs [28, 29].

This doubtful scenario of so far used MWNTs characterizing tools necessitates the identification of an efficient confirmatory tool, which not only produces reproducible results, but accurately recognizes MWNTs-content in an arc-generated carbonaceous sample as well, from a macroscopic point of view.

The science of composites containing MWNTs as filler is known for many years [30]. A number of reports [31-35] highlighting the percolation based ac conductivity enhancement of MWNT based dielectric composites are available in the literature. However, this branch of composite-science has never been used as a tool to characterize MWNT samples. The present study proposes the ac conductivity measurement of the composite materials, where the carbonaceous samples are impregnated at a well-anticipated filler loading in a dielectric host, as a reliable and confirmatory tool to characterize arc-synthesized MWNT-samples. The current proposition has been validated through a large number of measurements.

In order to demonstrate the ac conductivity as a characterizing tool, we chose to work with different batches of MWNTs with varying impurity (non-MWNT species) and structural disorders.

MWNT-samples employed in the present study were synthesized using a modified electric arc method, where both the purity and structural disorders of the as synthesized MWNTs were systematically controlled by an external focusing electrostatic voltage ($\Phi$) [17]. The synthesis methodology, TEM, TG and RS features of the as-synthesized MWNTs along with their plausible interpretations have been reported in our earlier communication [17]. Moreover, the first time report [25] of $\Phi$ by the authors showing its efficacy and the action mechanism on the formation of such MWNTs is also available for reference.

The ac conductivity responses of the as-synthesized MWNT samples and 99.99% pure graphite powder (Aldrich) were carried out by fabricating their composites with the 99.999% (Kemphasol) pure polystyrene (PS). Fabrication of the composite materials in the present study was carried out following the protocol available in the literature [36]. The as prepared composite films (20mm diameter and 0.2mm thickness) were coated with silver paste on both the surfaces before analyzing their electrical responses with the help of an impedance spectrometer (Navacontrol Technologies, alpha A High Performance Analyzer) over a frequency range of 0.1 Hz to10MHz. All the measurements were carried out at room temperature.

Figure 1 shows the results of the ac bulk electrical conductivity $\sigma$ for all the composites as a function of ac frequency ($\upsilon$).

Firstly, it is interesting to note that the value of $\sigma$ (0.8 Scm$^{-1}$) for the sample corresponding to $\Phi$=400V is about fourteen orders of magnitude higher than that of the virgin PS at the low frequency limit. Moreover, it is also seen that $\sigma$ shows no frequency dependent dispersal for the composites, whose fillers correspond to 0V≤$\Phi$≤400V. On the other hand, remaining composites show certain dependency on $\upsilon$ including the virgin PS host. It is also noteworthy that the value of $\sigma$ at each value of $\upsilon$ first increases when the filler corresponding $\Phi$ value is increased from 0V to 400V and then decreases monotonously up to 1200V; the trend being exactly similar to those exhibited by RS and TG measurements reported in our earlier study [17].

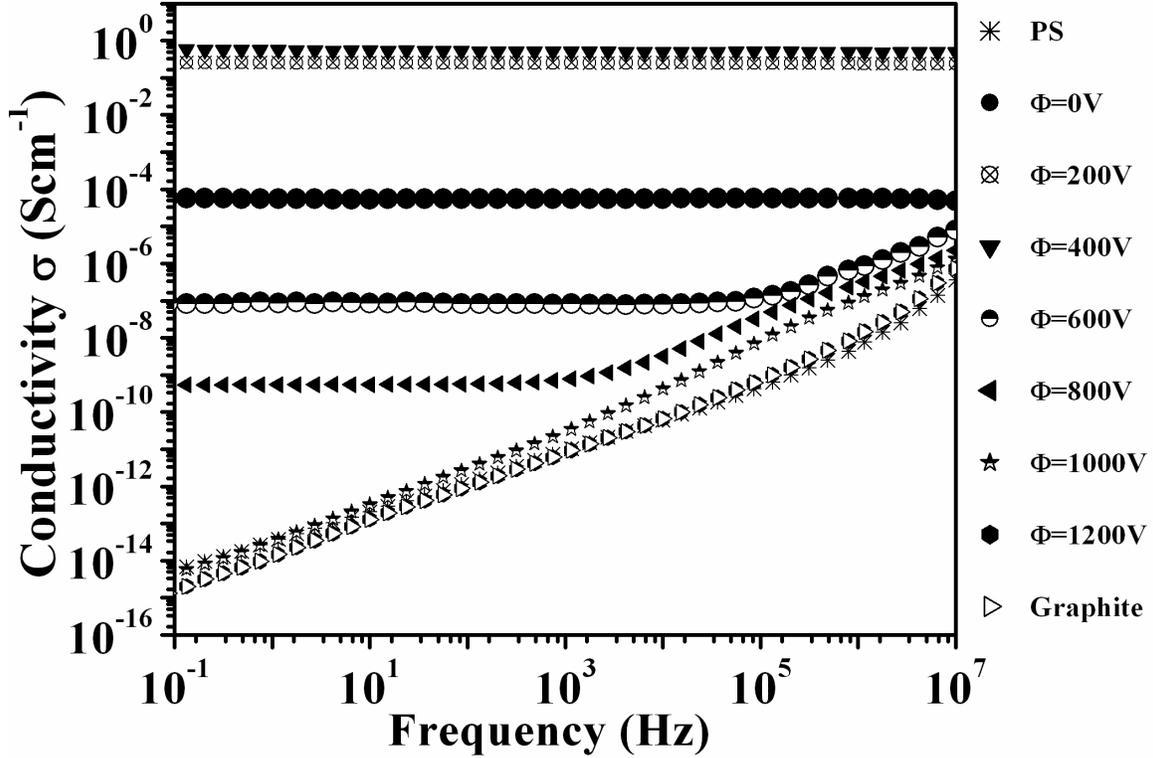

**Figure 1.** Variation of ac conductivity of the PS composites having different batches of MWNTs and fine graphite powder as filler along with the response of the virgin PS. The filler loading is fixed at 10wt% for all the composites and all the measurements were carried out at room temperature.

However, most interestingly, at the low $v$ limit, the graphite-PS composite shows a value of σ ($10^{-15}$ Scm$^{-1}$), which is about fourteen orders of magnitude lesser than that of the composite corresponding to Φ=400V.

This fourteen orders of magnitude difference in the ac conductivity value practically provides the identification bench-marks for both the graphite structure and the MWNTs. While analyzing both the RS and TG data [17], minor differences are seen for the MWNT-samples corresponding to Φ<400V and Φ>400V. However, the qualities of the samples are revealed much more clearly while looking at their electrical conductivity data (figure 1).

While both MWNTs and graphite are known for their outstanding electrical conductivity [37], later is much weaker in establishing a percolation network when impregnated in a dielectric host. Due to the very high aspect ratio, MWNTs are known to build up an appropriate percolation network showing an enhanced value of σ [36]. Moreover, CNPs, due to their spherical or oboid structures, are also of similar strength, like graphite, while taking part in the percolation network.

In the conventional arc assisted technique, cathode deposits are known [38] to contain nearly 60% MWNTs by weight; the rest being CNPs and amorphous carbon. It is noteworthy that this purity figure has been cited on account of direct weight measurements by separating the

MWNTs and CNPs by a non-destructive technique. Moreover, the percolation threshold for the composites with *extremely pure* MWNTs has been estimated to be roughly of the order of 8% [39, 40]. For our proposed tool to be successfully employed, the quantity of pristine nanotubes used as filler should not exceed 8% by weight. The present experiment used 10% by weight of the *as-prepared* MWNTs (with purity believed to be typically < 60%) as filler, which is well within the percolation threshold.

In figure 1, in going from the sample corresponding to Φ=0V to 400V the value of σ is seen to increase. Since, below the percolation threshold, the value of σ is expected to increase upon increasing MWNTs/CNPs ratio; ac conductivity in figure 1 predicts that, in the filler, MWNTs/CNPs ratio has increased on increasing Φ from 0V to 400V exhibiting a maximum at 400V. In figure 1, the value of σ at Φ=400V shows an excellent matching with that reported for a high-purity MWNT filler [36] and strongly supports our previous prediction that the MWNT-sample synthesized at Φ=400V is of very high purity. The above observations also validate our earlier conclusions while describing the effects of Φ in the range of 0V to 400V [17].

As Φ is increased beyond 400V, from figure 1 it is seen that σ shows an incremental tendency with frequency. Here, in moving from Φ=400V to 1200V, two observations are important: (i) the frequency lower limit, from which this incremental tendency in σ is observed, gradually shifts to the lower frequency side; (ii) at the low limit of $v$ the value of σ decreases monotonously and then finally overlaps with that of the PS matrix.

In order to infer the plausible rerasons for these observations, the intrinsic nature of the virgin PS host is worth paying attention. From the study of Dyre *et al* [41] it is known that any disordered material, irrespective of their intrinsic material properties, exhibits a universal ac conductive nature. Their study reveals that ac conductivity of a perfectly disordered solid strongly depends on the value of the frequency. Right from the low limit of $v$, the ac conductivity of all these materials exhibits frequency dependent dispersal. Our data for the virgin PS (figure 1), which is a well-known perfectly disordered solid, strongly support their claims.

In view of the above arguments, we strongly believe that those composites, which show almost no frequency dispersal, are dominated by the highly ordered crystalline structures. Arc-generated MWNTs are the bests among all the tubular graphene families known [6]. Our data, for the composites corresponding to 0V≤Φ≤400V, therefore, strongly reflect that the MWNTs used in the present work has an arc origin. This finding has its reliance if we compare our data with those of the composites having *pure* chemical vapor deposited MWNTs [42] as filler. Chemical vapour deposited MWNTs are known to contain certain amounts of defects and disorders in them [6] and this is why, even at a much higher filler concentration than the one used in the present study, they show frequency dependent dispersal [42].

Looking at the frequency dependent conductivity for the composites corresponding to Φ>400V and graphite powder as filler the following conclusions can readily be surmised.
(i) These composites are dominated by the defects and disorders present within them. The defect content increases monotonously as Φ is increased beyond 400V.
(ii) On increasing Φ beyond 400V, either the filler MWNTs correspond to defective ones, or CNPs/MWNTs ratio is higher in the corresponding fillers. On increasing CNPs/MWNTs ratio, the percolation network offered by the MWNTs is hampered. As a result, the intrinsic behavior of the PS base dominates. It can, therefore, be concluded that on increasing Φ beyond 400V, there is a monotonous increase in either defects in the filler MWNTs or MWNTs/CNPs ratio. This conclusion is also supported by our earlier hypothesis [17].

It is mention worthy that, in order to strengthen our present proposition, randomly selected five different composite samples cut from each composite film were tested separately under the present study and no significant variation in the value and nature of σ was observed. This finding is certainly an indication of the negligible percentage error while adopting our present protocol for characterizing MWNT-samples. However, more systematic investigations are required further to standardize this tool for routinely characterizing MWNT-samples of unknown and uncertain origins.

In conclusion, our study brings the fact to the limelight that ac electrical conductivity measurement is a very efficient, simple, cost-effective and confirmatory characterizing tool, which not only can distinguish between graphite and MWNTs, but can also reflect the reliable comparisons of the different batches of arc-grown MWNT-samples with respect to their overall quality and relative purity unambiguously. The large differences in the magnitudes of the ac conductivity data near the low frequency limit helps to increase the sensitivity of this technique over others. For this tool to be successfully employed, the only job one has to perform will be to fabricate composites using different *unknown* batches of MWNTs impregnated into a polymer matrix with a *fixed* filler loading lesser than the percolation threshold of the *pure* MWNTs and then analyze their ac electrical conductivity data. Our findings will definitely be important and significant from a realistic point of view if employed for understanding the exact roles of the controlling parameters and diverse methodologies involved in the former studies [8-16] (which have highlighted the enhancement of the overall yield and purity of MWNTs produced by arc evaporation) and many more studies to come up next.

**Acknowledgements**
This work was supported by the BRNS, DAE under the BARC-PU Joint Collaborative Research Program. HN acknowledges DRDO, Jodhpur for the financial support.